\documentclass[a4paper,11pt]{article}
\pdfoutput=1 
\usepackage{jcappub}

\usepackage[T1]{fontenc}

\usepackage{subfiles}
\usepackage{svg}
\usepackage{float}
\usepackage{hyperref}
\usepackage{multirow}
\usepackage{natbib}
\usepackage{printlen}
\usepackage{tikz}
\usetikzlibrary{positioning, arrows.meta, shapes.geometric, calc}

\definecolor{mypurple}{RGB}{164,64,214}
\definecolor{rarniblue}{RGB}{0, 0, 255}

\title{\boldmath Using Deep Learning for Robust Classification of Fast Radio Bursts}

\author[a]{Rohan Arni}
\author[b,c,d]{Carlos Blanco}
\author[c,e]{Anirudh Prabhu}

\affiliation[a]{High Technology High School, Lincroft NJ 07738}
\affiliation[b]{Department of Physics, The Pennsylvania State University, PA 16802, USA}
\affiliation[c]{Department of Physics, Princeton University, Princeton, NJ 08544, USA}
\affiliation[d]{Stockholm University and The Oskar Klein Centre for Cosmoparticle Physics,  Alba Nova, 10691 Stockholm, Sweden}
\affiliation[e]{Princeton Center for Theoretical Science, Princeton University, Princeton, NJ 08544, USA}

\emailAdd{roarni@ctemc.org}
\emailAdd{carlosblanco2718@princeton.edu}
\emailAdd{prabhu@princeton.edu}

\abstract{While the nature of fast radio bursts (FRBs) remains unknown, population-level analyses can elucidate underlying structure in these signals. In this study, we employ deep learning methods to both classify FRBs and analyze structural patterns in the latent space learned from the first CHIME catalog. We adopt a Supervised Variational Autoencoder (sVAE) architecture which combines the representational learning capabilities of Variational Autoencoders (VAEs) with a supervised classification task, thereby improving both classification performance and the interpretability of the latent space. We construct a learned latent space in which we perform further dimensionality reduction to find underlying structure in the data. Our results demonstrate that the sVAE model achieves high classification accuracy for FRB repeaters and reveals separation between repeater and non-repeater populations. Upon further analysis of the latent space, we observe that dispersion measure excess, spectral index, and spectral running are the dominant features distinguishing repeaters from non-repeaters. We also identify four non-repeating FRBs as repeater candidates, two of which have been independently flagged in previous studies.
}

\begin{document}

\maketitle
\flushbottom

\section{Introduction}
\label{sec:intro}

A major outstanding problem in contemporary astrophysics concerns the origin of fast radio bursts (FRBs). FRBs are bright ($\sim$Jy), short-duration ($\mu$s–ms) bursts of radio emission that typically originate from extragalactic or cosmological distances (see~\cite{Petroff:2021wug} for a recent review). Their burst rates exhibit significant source-to-source variability: some sources have produced only a single detected burst (“non-repeaters”), while others emit multiple bursts from the same location (“repeaters”)~\cite{Spitler:2016dmz, CHIMEFRB:2023myn}. Recently, a new class of “hyperactive repeaters” has been identified, with extraordinarily high activity--one source produced nearly four hundred bursts within an hour~\cite{Lanman:2021yba, Kirsten:2023eqd, 2022ATel15679....1M, Zhang:2023eui}. Despite this wealth of observations, the progenitors and emission mechanisms of FRBs remain poorly understood (see~\cite{Zhang:2022uzl} for a review of the observational data, radiation mechanisms, and proposed progenitors). One notable exception is FRB 200428, which was detected in coincidence with an X-ray burst from a Galactic magnetar, suggesting a possible connection between (at least some) FRBs and magnetars~\cite{Bochenek:2020zxn, CHIMEFRB:2020abu}.

The diversity in observed FRB burst rates suggests the possibility of multiple progenitor channels. A natural hypothesis is that repeating and apparently non-repeating FRBs correspond to distinct source classes, with the latter arising from catastrophic events. However, this scenario is generally disfavored because the FRB event rate exceeds that of known cataclysmic processes~\cite{Ravi:2019iop}. Even if all FRBs ultimately repeat, an open question is whether multiple classes of repeaters exist. Based on the current sample, Ref.~\cite{Zhang:2022uzl} identifies three observational categories: (1) regularly active repeaters, (2) less energetic and less regular repeaters, with currently “non-repeating’’ sources potentially belonging to this group, and (3) highly active yet low-luminosity repeaters, exemplified by the globular-cluster source FRB 20200120E~\cite{Nimmo2022}. Distinguishing among progenitor models therefore requires robust classification of FRBs as repeaters or non-repeaters, as well as more general clustering schemes, since non-detection of bursts does not guarantee true non-repetition. Sources may simply emit below telescope sensitivity or with long waiting times. The situation is further complicated by sources like FRB 200428, which exhibit multiple sub-bursts~\cite{Bochenek:2020zxn, CHIMEFRB:2020abu} and blur simple classification boundaries.

Several approaches have been proposed to address this classification issue. Observationally, there are indications of systematic differences between the bursts of repeaters and non-repeaters. Compared to non-repeaters, repeating sources tend to exhibit more complex, multi-component temporal structures (e.g., downward-drifting subpulses), narrower spectra, higher rotation measures, and greater variability~\cite{CHIMEFRB:2019pgo}. However, no single observational criterion has yet been established to robustly distinguish repeaters from non-repeaters. Another approach that has been explored involves machine-learning-based classification. A variety of machine-learning methods have been developed to classify FRBs into repeaters and non-repeaters. Supervised machine-learning algorithms, including random forests, logistic regression, gradient boosting, and shallow neural networks, have been applied to FRB catalog datasets~\citep{Luo:2022smj, Zhu-Ge:2022nkz, J_nior_2026}. These approaches give reasonably high accuracy but lack interpretability of results, and cannot find underlying patterns in the data effectively. Variational Autoencoders (VAEs) are a class of machine learning models that learn compact representations of high-dimensional data~\citep{VAEpaper}. By encoding data into a latent space governed by a Gaussian distribution, VAEs offer a dimensionality reduction method while preserving important features. Supervised Variational Autoencoders (sVAEs) extend this approach by incorporating a classification objective alongside the representation learning task~\citep{svaepaper}. This dual-task framework not only improves the classification performance but also enhances the interpretability of the latent space by revealing potential patterns in the data. Generative models have shown the ability to learn compact, continuous latent distributions of high-dimensional data~\citep{VAEpaper}. Combining a generative model such as a Variational Autoencoder along with a classification objective allows us to utilize both the descriptive and discriminative abilities of neural networks, along with allowing for more interpretability of results. This paper aims to utilize an sVAE to classify FRBs as repeaters or non-repeaters while analyzing the learned latent space for meaningful structures and outliers. By optimizing the model for both classification accuracy and latent space regularization, we attempt to advance the state-of-the-art in FRB classification and indicate learned patterns in the sVAE latent space. 

The organization of the paper is as follows. In Sec.~\ref{sec:methods}, we describe the model for latent space embedding, clustering, and classification. This section also outlines evaluation metrics for the classification performance. The results of the latent space analysis are presented in Sec.~\ref{sec:results}. An interpretation of the results in the context of fast radio burst phenomenology, along with future directions, is outlined in Sec.~\ref{sec:discussion}.

\section{Methods}
\label{sec:methods}

Here, we describe our model for embedding, clustering, and classifying FRB data. We construct a model that aims to (1) embed the discrete space of FRB signals into a continuous latent space, and (2) classify or cluster the signals in order to predict the potential for signal repetition. By embedding the FRB data into a continuous latent space, we can take advantage of a well-defined similarity measure--Euclidean distance in the latent space-- as a natural method to find substructure. We adopt a machine-learning architecture based on two well-studied frameworks, the VAE, and the multi-layered perceptron (MLP). 

An MLP is a simple artificial neural network of fully-connected layers, characterized by its input and output dimensions, as well as its depth (number of layers) and its hidden dimension (the dimension of all non-terminal layers). Since MLPs excel at regression and classification tasks, we use this architecture in order to predict whether or not an FRB signal is associated with a repeater source. The training is done to maximize the accuracy of classification using the known labels of the FRB data. 

A VAE is another kind of neural architecture where the dimension of the layers is reduced from the input dimension of the data down to a bottleneck characterized by a latent dimension. This initial constriction, so-called \emph{encoding}, can be thought of as a compression of information from a discontinuous data space into a lower-dimensional \emph{latent space}. Importantly, a VAE represents data points as multivariate Gaussian distributions in latent space, i.e. each data point is mapped into the latent space as a multivariate Gaussian characterized by a mean $\mu$ and a variance $\sigma$. The second half of the VAE, the \emph{decoder}, is another fully-connected network that expands the dimensionality of the layers back into the dimension of the data space. The training of a VAE minimizes the loss of information as a data point is passed through the network by attempting to make the output of the VAE as close to the input as possible. In doing so, the VAE implicitly constructs a latent representation of the data set, via a minimally-lossy compression. Since the latent space is a smooth and simply-connected space of probability distributions, the data can be represented by regions rather than points.

A supervised VAE (sVAE) combines the regressive strengths of an MLP with the generative and embedding strengths of a VAE. The idea is to construct an information-dense latent representation of the FRB data from which to learn substructure and classify the signals. The VAE is trained to have the highest fidelity in encoding and decoding our data, therefore generating a latent embedding. The MLP takes a point in the latent space as an input and predicts the probability that the data point is a repeater. The training is done concurrently in order for the classification of the data to inform the structure of the latent space.

\subsection{Data Preprocessing}

Our primary data source is the first CHIME (Canadian Hydrogen Intensity Mapping Experiment) FRB catalog~\cite{CHIMEFRB:2021srp}, which contains 536 unique FRB sources that produced 570 detected bursts, including both repeating and non-repeating sources (see Table~\ref{tab:class_distribution}). We use data from a single telescope to minimize potential instrument-dependent selection effects, such as biases introduced by differences in time resolution that could affect burst width measurements. Each data point represents the features from a single burst.

\begin{table}[H]
\centering
\begin{tabular}{|l|c|c|}
\hline
\textbf{Class} & \textbf{Count} & \textbf{Percentage} \\
\hline
Non-Repeaters & 476 & 83.5\% \\
\hline
Repeaters & 94 & 16.5\% \\
\hline
\end{tabular}
\caption{Class distribution in the FRB dataset from the first CHIME/FRB catalog~\cite{CHIMEFRB:2021srp}.}
\label{tab:class_distribution}
\end{table}

From this data, we selected features that describe each signal, but do not divulge the (non-)repeating nature of the source (see Table~\ref{tab:features}). For features with quoted measurement uncertainties, we created separate upper and lower bound features by adding and subtracting the reported errors. The dataset was standardized using \texttt{StandardScaler} from the \texttt{sklearn} library~\citep{sklearnpaper}. 

The dispersion measure (DM) is the integrated column density of free electrons along the path of an FRB. It is determined by analyzing dynamic spectra using the Bonsai and fitburst algorithms from CHIME, which search for the optimal DM that de-disperses the signal~\cite{chimeproceedingpaper}. The flux, measured in units of Janskys (Jy), is the power detected per unit area per unit frequency, and the fluence, measured in units of Jy ms, is the flux integrated over the duration of the burst.  The frequencies characterize the frequency band observed from the burst, and the spectral index and running characterize the flux spectrum. The DM excess describes the DM minus the Milky Way contribution, computed using the NE2001~\cite{NE2001} and YMW16~\cite{Yao_2017} models of galactic electron density distribution.

\begin{table}[h]
\centering
\begin{tabular}{|l|l|}
\hline
\textbf{Name} & \textbf{Description} \\
\hline
DM (fitburst)  & DM determined using the fitting algorithm fitburst \\ \hline
Fluence & Fluence of burst \\ \hline 
Flux & Flux of burst \\ \hline
Freq (high) & Highest frequency of the observation \\ \hline
Freq (low) & Lowest frequency of the observation \\ \hline
Freq (peak) & Center frequency of the observation \\ \hline
Spectral Index & Spectral index for the sub-burst \\ \hline
Spectral Running & Spectral running for the sub-burst \\ \hline
DM (Bonsai) & DM determined using the Bonsai detection algorithm \\ \hline
DM excess (NE2001) & DM excess between DM determined by fitburst and NE2001  \\ \hline
DM excess (YMW16) & DM excess between DM determined by fitburst and YMW16  \\ \hline
Width & Box car width of the pulse \\ \hline
\end{tabular}
\caption{\label{tab:features} Descriptions of selected features used in the dataset.}
\end{table}

We discarded date-time information (MJD labels), as these features could leak the repeating nature of sources, as multiple bursts from the same source would have correlated arrival times. We also discarded burst localization features (galactic longitude/latitude, sky position) and quality metrics (SNR values, chi-squared, degrees of freedom), as the former could allow identification of known repeating sources while the latter are detector-specific and not related to the nature of each FRB.

\subsection{Model Architecture}

\begin{figure}[H]
    \centering
    \includegraphics[width=0.8\textwidth, trim=1cm 2cm 1cm 2cm, clip]{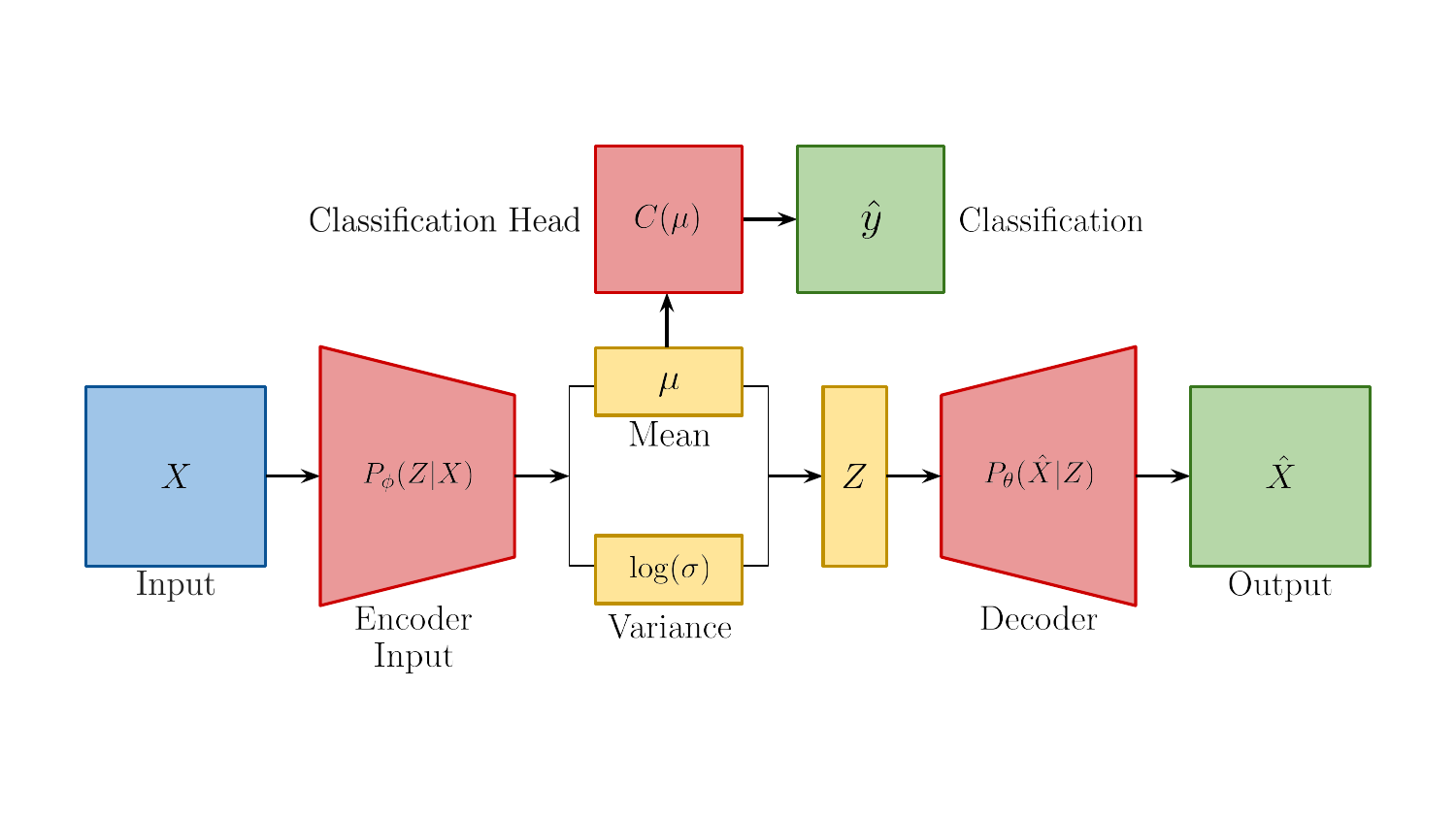}

    \caption{Schematic of model architecture.}
    \label{fig:i}
\end{figure}

In this section, we describe our model architecture in detail. The sVAE consists of three main components: an encoder, a decoder, and a classification head~\citep{svaepaper}. The encoder compresses the input data into a lower-dimensional latent representation, while the decoder reconstructs the input from the latent space. The classification head simultaneously predicts whether an FRB is a repeater or non-repeater. Below we describe each of these components in more detail. 

\subsubsection*{Encoder}

We parameterize the encoder as a neural network with trainable parameters $\phi$, which learns to map input features $X$ to the parameters of a latent Gaussian distribution. The encoder network $P_\phi(\mu, \log \sigma | X)$ is composed of three fully-connected layers, and compresses input features $X$ into features characterizing a normally-distributed lower-dimensional latent space. Each layer is followed by batch normalization, which normalizes the activations of each layer across a mini-batch, and dropout regularization, which randomly sets a fraction of neurons to zero during training. These modifications help training stabilization~\citep{JMLR:v15:srivastava14a, ioffe2015batchnormalizationacceleratingdeep}. For each input vector, the network outputs two lower-dimensional vectors: one corresponding to the mean $\mu$ and the other to the log variance $\log \sigma ^2$ of the latent Gaussian distribution.

\subsubsection*{Latent Space}
The latent space is a compressed, probabilistic representation of the input FRB data. This latent space is constrained to follow a multivariate Gaussian distribution with mean $\mu$ and diagonal covariance matrix determined by $\sigma^2$. This probabilistic formulation enables the model to learn a continuous, structured embedding where similar FRB signals are positioned close to one another. This  latent space becomes valuable for data analysis, as it allows for clustering, outlier detection, and visualization of FRB population structure through dimensionality reduction techniques. This latent space is also structured for interpretability of the classifier head.

If we try to train a model that samples the latent space as $Z \sim \mathcal{N}(\mu, \sigma^2)$, we cannot compute the gradient of the loss function through backpropagation during the training process, as stochastic sampling is a non-differentiable process. To enable backpropagation through the stochastic latent variables, we apply the reparameterization trick. We draw a noise vector from the standard normal distribution $\epsilon \sim \mathcal{N}(0, I)$  and compute 
\begin{align}
    Z &= \mu + \sigma \cdot \epsilon \\
    \sigma &= \exp\left(\frac{1}{2} \log \sigma^2\right)
\end{align}
where $\mu$ and $\log \sigma^2$ are the mean and log variance vectors of the latent space, and $Z$ is our reparametrized vector. By computing this expression, the source of randomness in $Z$ is independent of $\mu$ and $\sigma$ given by the neural network, allowing us to backpropagate through the random sampling process.

\subsubsection*{Decoder}
The decoder is parameterized as a neural network with trainable parameters $\theta$, which reconstructs the original input $\hat{X}$ from the latent representation $Z$. The decoder network $P_\theta(\hat{X}| Z)$ mirrors the encoder structure and is responsible for reconstructing the input from the latent representation. It consists of three fully-connected layers with batch normalization and dropout regularization and takes in the sampled latent vector $z$ and maps it to the original feature space. The final layer outputs a vector of the same dimensionality as the original input. 

\subsubsection*{Classification Head}

The classification head is designed for binary classification (repeater vs. non-repeater). It consists of three fully-connected layers, and takes the $\mu$ vector as input and outputs a single logit representing classification.

\subsection{Training Procedure}

We trained the Supervised VAE using a stratified 5-fold cross-validation scheme to evaluate performance on the full dataset without data leakage.  The dataset was partitioned into five folds, each preserving the repeater vs. non-repeater ratio. For each fold, four parts were used for training and the remaining part for validation; this process was repeated until each fold had served as the validation set exactly once.  Within each fold, data was loaded in mini-batches of 64.

We optimized all model parameters with the Adam optimizer~\citep{kingma2017adammethodstochasticoptimization}. We applied a learning rate scheduler to reduce the learning rate by a factor of 2 whenever the validation loss plateaued. To prevent overfitting, we employed an early stopping condition on the validation loss.

The model optimizes a combination of three loss components. The reconstruction loss measures the discrepancy between the reconstructed and original input using Mean Squared Error (MSE).
The KL Divergence regularizes the latent space to follow a Gaussian distribution, written as:
    \begin{equation}
        \mathcal{L}_{\rm KL} = -\frac{1}{2} \sum_{i=1}^{d} (1 + \log \sigma^2_i - \mu_i^2 - \sigma^2_i),
    \end{equation}
where $i$ is the latent feature index, $d$ is the total number of features, $\sigma_i^2$ is the variance of that latent feature, and $\mu_i$ is the mean of that latent feature.

Finally, the classification loss uses Binary Cross-Entropy with logits to predict FRB classes. A class weight is applied to balance the dataset. During each training epoch, we accumulated per‑batch reconstruction, KL, and classification losses, backpropagated the total loss, and updated all network weights. The parameters for both the classification head and the VAE are updated simultaneously in each backpropagation step. The completed, weighted loss function for the training loop is defined as:

\begin{equation}
    \label{eq:lossfn}
\mathcal{L}_{\rm total} \;=\;\mathcal{L}_{\rm recon}\;+\;\beta\,\mathcal{L}_{\rm KL}\;+\;\gamma\,\mathcal{L}_{\rm class}\,,
\end{equation}
where $\beta$ and $\gamma$ weight the KL divergence and classification losses, respectively. The choice of these parameters is discussed in the following section.

We optimized the model hyperparameters using Gaussian Process models with the Optuna library~\cite{optunapaper}. This is faster than a brute-force grid search, which is typically the standard hyperparameter optimization process. More details are in Section~\ref{sec:hyp_opt_results}.

\subsection{Evaluation Metrics}
\label{sec:metrics}

We evaluate classification performance using weighted accuracy, precision, recall, F1 score, and F2 score~\citep{sklearnpaper}. Let $C$ be the set of classes (repeater or non-repeater), and $w_c$ be the weight of class $c$, typically defined as the proportion of true instances in that class. Then $\mathrm{TP}_c$, $\mathrm{FP}_c$, and $\mathrm{FN}_c$ denote the true positives, false positives, and false negatives for class $c$, respectively. We compute our performance metrics as follows,

\begin{equation}
\mathrm{Accuracy} = \sum_{c \in C} w_c \cdot \frac{\mathrm{TP}_c + \mathrm{TN}_c}{\mathrm{TP}_c + \mathrm{TN}_c + \mathrm{FP}_c + \mathrm{FN}_c},
\end{equation}

\begin{equation}
\mathrm{Precision} = \sum_{c \in C} w_c \cdot \frac{\mathrm{TP}_c}{\mathrm{TP}_c + \mathrm{FP}_c},
\end{equation}

\begin{equation}
\mathrm{Recall} = \sum_{c \in C} w_c \cdot \frac{\mathrm{TP}_c}{\mathrm{TP}_c + \mathrm{FN}_c},\end{equation}

\begin{equation}
\mathrm{F_1} = \sum_{c \in C} w_c \cdot 2 \cdot \frac{\mathrm{Precision}_c \cdot \mathrm{Recall}_c}{\mathrm{Precision}_c + \mathrm{Recall}_c},
\end{equation}

\begin{equation}
\mathrm{F_2} = \sum_{c \in C} w_c \cdot 5 \cdot \frac{\mathrm{Precision}_c \cdot \mathrm{Recall}_c}{4 \cdot \mathrm{Precision}_c + \mathrm{Recall}_c}.
\end{equation}

\subsection{Latent Space Analysis}
\label{sec:methods_and_analysis}

After training the sVAE, we examined the structure of the latent space to uncover potential patterns or groupings relevant to FRB classification. The $\mu$ vectors produced by the encoder represent the learned latent embeddings of input samples. 

To visualize the high-dimensional latent space, we used two dimensionality reduction techniques. Principal Component Analysis derives a linear transformation that captures the directions of maximum variance in the data~\citep{hotelling1933analysis}. Another dimensionality reduction technique is t-distributed Stochastic Neighbor Embedding (t-SNE) which provides a non-linear embedding focused on preserving local structures in the higher-dimensional data~\citep{tsnemaatenhinton}.

We compute the Pearson and Spearman correlation between input features and reduced latent space features (PCA \& t-SNE features) as a preliminary metric to understand the composition of latent features~\citep{a2018_correlation}.

Furthermore, we perform latent space analysis to see how one changing individual latent feature affects repeater probability and physical features while ignoring the other latent features; which we call partial dependence analysis ~\citep{Molnar_2025}. We also generate partial dependence plots to visualize how physical features and repeater probability evolve based on one feature of the latent space. For each latent feature $z_s$, we can examine how traversing that feature affects both the input features and repeater probability. We can traverse the latent space over a characteristic range based on the expected distribution from minimizing KL divergence.

\section{Results} \label{sec:results}

Here, we present our main results, show the final optimized hyperparameters of the model, and analyze the classification performance based on the benchmarks outlined in Section~\ref{sec:metrics}. Furthermore, we identify possible repeater candidates based on false positives, and perform the latent space analysis detailed in Section~\ref{sec:latent_pdp}.

\subsection{Optimized Hyperparameters}
\label{sec:hyp_opt_results}

Our model had a set of hyperparameters that must be optimized separately from the training procedure, as they are parameters that determine the architecture of the model. The hidden and latent dimensions are the size of the encoder/decoder layers and the latent space, respectively. The KL weight and classification weight were used to scale the components of the loss function, as shown in Equation~\ref{eq:lossfn}. The dropout rate determined the probability of randomly setting layer outputs to zero during training to prevent overfitting, and the learning rate controlled the step size of the gradient descent optimizer during backpropagation. The scheduler patience was used to determine how many epochs the model training would go before lowering the learning rate. The classification multiplier was a secondary scale factor for the classification loss to ensure that it would not be outweighed by the other components of the loss function, as classification loss is orders of magnitude smaller than reconstruction loss and KL loss. The activation function is the non-linear function applied at the end of each layer in the sVAE. 

The model was optimized using an Optuna hyperparameter search over 350 trials, using a stratified 5-fold cross-validation to evaluate each trial. After identifying the best trial based on classification head accuracy, we retrained the model on the full training set for up to 150 epochs, again with early stopping on validation loss. We present the optimal hyperparameters in Table~\ref{tab:hyperparams}. 

\begin{table}[H]
\centering
\begin{tabular}{|l|l|}
\hline
\textbf{Hyperparameter}            & \textbf{Value}                          \\ \hline
Hidden dimension (\texttt{hidden\_dim})   & 1530                                    \\ \hline
Latent dimension (\texttt{latent\_dim})   & 16                                      \\ \hline
KL weight (\(\beta\))              & 1.2212                                  \\ \hline
Classification weight (\(\gamma\)) & 0.5886                                  \\ \hline
Dropout rate                       & 0.1097                                  \\ \hline
Learning rate (\texttt{lr})              & 0.0001308                               \\ \hline
Scheduler patience                 & 7                                       \\ \hline
Positive class weight              & 0.8946                                  \\ \hline
Activation function                & ReLU                                    \\ \hline
Classification multiplier          & 12452.1433                               \\ \hline
\end{tabular}
\caption{Final hyperparameter values for the optimized sVAE model, after 350 Optuna trials.}
\label{tab:hyperparams}
\end{table}

\subsection{Optimized Model Layers}

The optimized sVAE architecture is detailed in Table~\ref{tab:architecture}. The encoder compresses 17 input features through three fully-connected (FC) layers (hidden dimension 1530) into a 16-dimensional latent space parameterized by $\mu$ and $\log \sigma^2$. The decoder reconstructs the input from latent samples using a symmetric architecture. The classification head processes the latent mean $\mu$ through three layers with progressively decreasing dimensions (765, 383, 1) to produce repeater probability predictions. Each layer incorporates ReLU activation, batch normalization, and dropout for training stability.

\begin{table}[H]
\centering
\begin{tabular}{|l|c|l|l|}
\hline
\textbf{Component} & \textbf{Layer} & \textbf{Dimensions} & \textbf{Operations} \\
\hline
\multirow{5}{*}{\textbf{Encoder}} 
& FC 1 & $17 \rightarrow 1530$ & ReLU, BatchNorm, Dropout \\
& FC 2 & $1530 \rightarrow 1530$ & ReLU, BatchNorm, Dropout \\
& FC 3 & $1530 \rightarrow 1530$ & ReLU, BatchNorm, Dropout \\
& $\mu$ & $1530 \rightarrow 16$ & \hfill --- \hfill \\
& $\log \sigma^2$ & $1530 \rightarrow 16$ & \hfill --- \hfill  \\
\hline
\multirow{4}{*}{\textbf{Decoder}} 
& FC 1 & $16 \rightarrow 1530$ & ReLU, BatchNorm, Dropout \\
& FC 2 & $1530 \rightarrow 1530$ & ReLU, BatchNorm, Dropout \\
& FC 3 & $1530 \rightarrow 1530$ & ReLU, BatchNorm, Dropout \\
& Output & $1530 \rightarrow 17$ & Linear \\
\hline
\multirow{3}{*}{\textbf{Classifier}} 
& FC 1 & $16 \rightarrow 765$ & ReLU, BatchNorm, Dropout \\
& FC 2 & $765 \rightarrow 383$ & ReLU, BatchNorm, Dropout \\
& Output & $383 \rightarrow 1$ & \hfill --- \hfill\\
\hline
\end{tabular}
\caption{Architecture of the SupervisedVAE model. The encoder compresses 17 input features to a 16-dimensional latent space characterized by mean $\mu$ and variance $\sigma^2$. The decoder reconstructs the original features from latent samples. The classifier predicts repeater probability from the latent mean.}
\label{tab:architecture}
\end{table}

As shown in Table~\ref{tab:model_size}, the complete model contains approximately 9.82 million parameters (37.47 MB). 

\begin{table}[H]
\centering
\begin{tabular}{|l|l|}
\hline
\textbf{Parameter}                        & \textbf{Value}          \\ \hline
Model parameters                         & 9,822,649               \\ \hline
Total memory size of trainable parameters & 37.47 MB                \\ \hline
\end{tabular}
\caption{Model size statistics for the optimized sVAE model.}
\label{tab:model_size}
\end{table}

\subsection{Classifier Performance}

The optimized sVAE was trained for 150 epochs on the final dataset. The model achieved strong performance on the FRB classification task. 
Tables \ref{tab:conf_matrix_signal} and \ref{tab:metrics_signal} summarize the signal-level confusion matrix and classification metrics.

\begin{table}[h]
\centering
\begin{tabular}{|l|c|c|}
\hline
 & \textbf{Predicted Non-Repeater} & \textbf{Predicted Repeater} \\
\hline
\textbf{True Non-Repeater} & 472 & 4 \\
\hline
\textbf{True Repeater} & 7 & 87 \\
\hline
\end{tabular}
\caption{\label{tab:conf_matrix_signal} Signal-level confusion matrix.}
\end{table}

\begin{table}[H]
\centering
\begin{tabular}{|lr|c|}
\hline
\textbf{Metric} & & \textbf{Value} \\
\hline
Accuracy & & 0.9807 \\
\hline
Precision & & 0.9805 \\
\hline
Recall (Sensitivity) & & 0.9807 \\
\hline
F1 Score & & 0.9806 \\ 
\hline
F2 Score & & 0.9807 \\ 
\hline
\end{tabular}
\caption{\label{tab:metrics_signal} Signal-level classification metrics.}
\end{table}

Overall, the sVAE exhibited high signal-level classification performance.

\subsection{False Positive Analysis}
\label{sec:fpanalysis}

The model had four false positives (cases where a non-repeater signal was flagged as a repeater). These false positives could indicate incorrectly classified signals, as non-repeaters have the potential to be unobserved repeaters. We present these FRB signals in Table~\ref{tab:false_positives}. 

\begin{table}[H]
\centering
\begin{tabular}{|l|c|}
\hline
\textbf{FRB Signal} & \textbf{Previously Proposed?}\\ 
\hline
FRB20181218C & Yes~\cite{Luo:2022smj} \\
\hline
FRB20190122C & No \\
\hline
FRB20190221A & Yes~\cite{Luo:2022smj, Zhu-Ge:2022nkz, J_nior_2026} \\
\hline
FRB20190320A & No \\
\hline
\end{tabular}
\caption{\label{tab:false_positives} Non-repeater FRB signals incorrectly classified as repeaters. These four false positives represent potential candidates for future repeater detection.}
\end{table}

Two of these signals, FRB20181218C and FRB20190221A, have been independently identified as repeater candidates in previous literature ~\cite{Zhu-Ge:2022nkz, Luo:2022smj, J_nior_2026}. Interestingly, these two FRBs are false positives from different methodological approaches. FRB20181218C was identified as a candidate by  a variety of supervised machine learning methods with a relatively low confidence score of 147 (out of 1000 trials on different data splits)~\cite{Luo:2022smj}. In addition, it was identified using a variety of clustering algorithms on FRB data, but needed additional features to appear through that approach~\citep{J_nior_2026}. In contrast, FRB20190221A appeared in all repeater candidate lists with high confidence~\cite{Zhu-Ge:2022nkz, Luo:2022smj, J_nior_2026}. The fact that our VAE independently identifies these same candidates strengthens the case for follow-up observations.

\subsection{Latent Space Visualization}

\begin{figure}[H]
  \centering
    \includegraphics[width=0.49\textwidth]{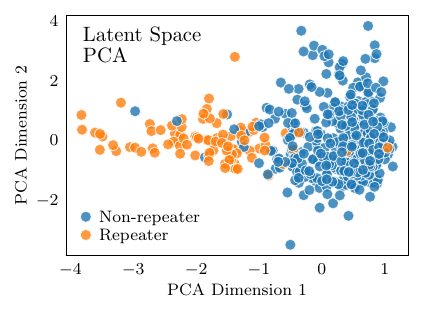} 
    \includegraphics[width=0.49\textwidth]{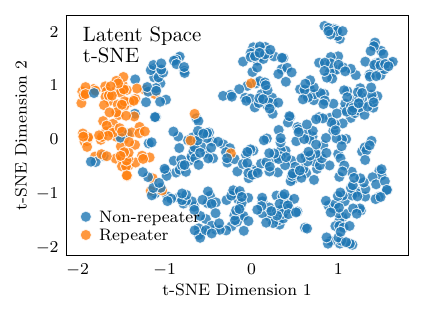}
  \caption{PCA and t-SNE of Latent Space Representations, Colored by Repeater Class.}
  \label{fig:full_tsne_frb}

\end{figure}

We can perform dimensionality reduction techniques (PCA, t-SNE) on the higher-dimensional latent space to visualize the latent data. As we show in Figure~\ref{fig:full_tsne_frb}, the latent space separates the data into repeaters and non-repeaters.

\subsection{Correlation Tests}

\begin{figure}[H]
  \centering

  \includegraphics[width=0.57\textwidth]{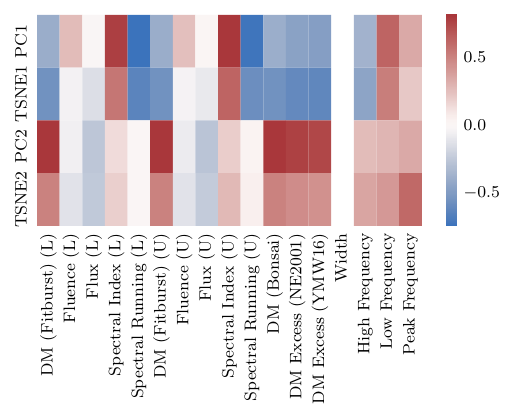}

  \caption{Spearman correlation heatmap between feature space and PCA (left) and t-SNE (right) axes.}
  \label{fig:correlation_projection}
  
\end{figure}

To identify what features are being visualized in each dimensionality reduction plot, we first compute the dimensionality reduction from the 16-dimensional latent space into a 2-dimensional ``reduced'' space through PCA or t-SNE. Subsequently, we compute the Spearman test between each ``reduced'' space and the feature space, which is visualized in Figure~\ref{fig:correlation_projection}.

We can see major similarities between both the PCA and the t-SNE feature correlations. This tells us that the choice of dimensionality reduction is somewhat irrelevant, and that the structures we observe in the plots of the latent space are not artifacts of a specific dimensionality reduction technique. We conclude that these patterns are representative of higher-dimensional structure within the latent space.

\subsection{Latent Space Partial Dependence Plots}
\label{sec:latent_pdp}

To examine how individual latent features affect probability, we can marginalize over all but one latent feature. We use the partial dependence function for classification~\cite{Molnar_2025}, which is defined as

\begin{align}
    \hat{f_s}(z_s) = \int \hat{f}(z_s, Z_C) d \mathbb{P}(Z_C)
\end{align}
where $\hat{f}$ is the classifier network, $z_s$ is the latent feature that we are interested in, and $Z_C$ are the latent space features we do not care about. We can approximate this as

\begin{align}
    \hat{f_s}(z_s) = \frac{1}{n} \sum_{i=1}^n\hat{f}(z_s, z_c^i),
\end{align}
where $z_c^i$ are the feature values from the dataset that we are not interested in, and $n$ is the number of latent vectors. The averages effectively provide a partial dependence function based on the empirical data that the model is trained on.

We can track how input features evolve as we traverse the latent space. For each latent feature, we choose the two most correlated input features based on the Spearman test. For a given physical feature $g$ of index $m$, we compute its average reconstructed value as:

\begin{align}
\label{eq:decoded_feat_evo}
\hat{g}(z_s) = \frac{1}{n} \sum_{i=1}^n [d(z_s, z_c^i)]_m,
\end{align}
where $d$ denotes the decoder network. Here, $[d(z_s, z_c^i)]_m$ is the $m$-th output component of the decoder when the latent vector is formed by concatenating the fixed value $z_s$ (for the latent dimension of interest) with $z_c^i$ (all other latent dimensions from data point $i$). This gives the average reconstructed feature value at each step along $z_s$. We quantified the impact of each latent dimension on classification by computing the marginalized repeater probability density function from $z_s \in [-3, 3]$. Since the KL divergence loss should construct a latent space which is close to a unit multivariate Gaussian, the range $z_s \in [-3, 3]$ approximately captures 3 standard deviations of the data. 

\begin{table}[h!]
\centering
\begin{tabular}{|c|c|c|}
\hline
Latent & Top Features & Probability Change \\
\hline
z13 & Spectral Running (L), Spectral Index (U) & 0.1895 \\ \hline
z9 & Spectral Running (U), Spectral Running (L) & 0.1754 \\ \hline
z7 & DM Excess (NE2001), DM Excess (YMW16) & 0.1745 \\ \hline
z15 & Spectral Index (U), Spectral Index (L) & 0.1667 \\ \hline
z3 & Low Frequency, Peak Frequency & 0.1370 \\ \hline
z4 & Spectral Index (U), Low Frequency & 0.1182 \\ \hline
z8 & Spectral Index (U), Low Frequency & 0.1108 \\ \hline
z6 & High Frequency, Peak Frequency & 0.1042 \\ \hline
\end{tabular}
\caption{Top features and corresponding probability change. We describe the lower and upper bounds of each feature as (L) and (U), respectively.}
\label{tab:latent_feature_probability_change}
\end{table}

\begin{figure}[H]
  \centering
  \includegraphics[width=1\textwidth]{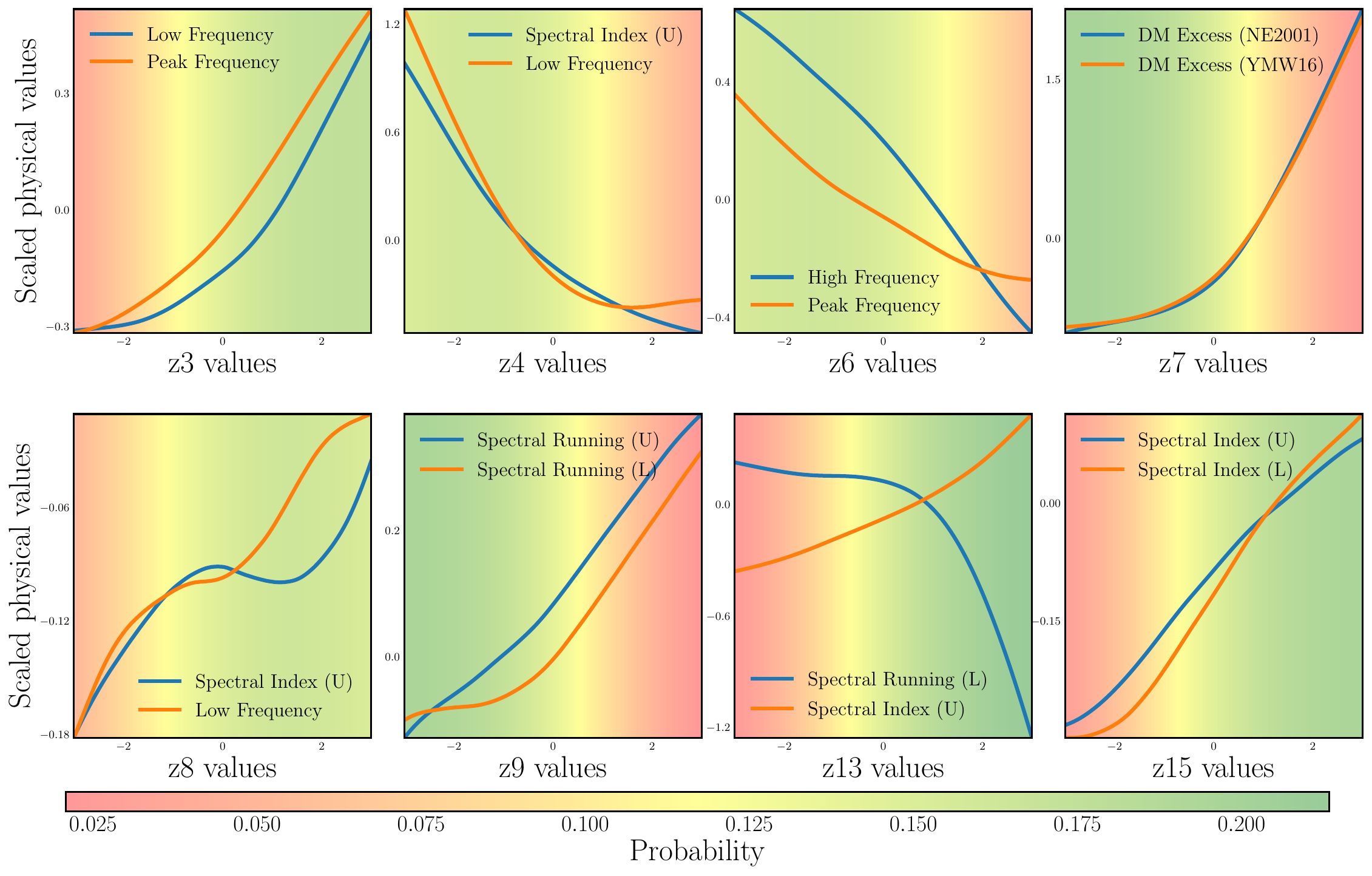}
  \caption{Latent space traversal for the most impactful features. The change in the top two correlated input features is shown. Repeater probability is shown as a background gradient.}
  \label{fig:latent_space_traversal}
\end{figure}

Figure~\ref{fig:latent_space_traversal} shows the partial dependence plots for latent features where the marginalized repeater probability distribution function has a range greater than $0.1$. These latent features are described in Table~\ref{tab:latent_feature_probability_change}. For each plot, the probability change is represented by the background color map. For each latent feature, the most correlated data features, given by the Spearman test, are plotted over the range $[-3, 3]$, as given by Eq.~\ref{eq:decoded_feat_evo}. 

\section{Conclusions and Discussion} \label{sec:discussion}

In this work, we applied a Supervised Variational Autoencoder (sVAE) to classify FRBs as repeaters or non-repeaters while simultaneously learning an interpretable latent space representation of the data. By combining classification with generative modeling, we created a model that is accurate in classification and also interpretable for more physical understanding and analysis. The sVAE achieved high signal-level accuracy, with an $\mathrm{F_2}$ score of $0.9807$, which is significantly higher than previous methods from~\cite{Luo:2022smj} and~\cite{Zhu-Ge:2022nkz}, with the highest $\mathrm{F_2}$ score being $0.8180$. In addition, the sVAE revealed clear structure in the learned latent space. The model was able to generate unsupervised clusters in the latent space, indicating the ability to group similar signals. Linear and non-linear embeddings of the latent space show distinction between repeaters and non-repeaters, which indicates separate progenitor models for repeaters and non-repeaters.

Another benefit of the sVAE is its improved interpretability: it allows us to identify key observables that are the best predictors of repeater status. 

From our latent space analysis (see Table~\ref{tab:latent_feature_probability_change}), repeater likelihood is most strongly associated with higher spectral index ($\gamma$) and lower spectral running ($r$). These parameters are obtained by fitting the spectrum of each burst to

\begin{align}
F(\gamma, r) = \left( {\frac{f} {f_0}} \right)^{\gamma + r \ln(f/f_0)},
\end{align}
where $F$ is the flux spectrum and $f_0 \approx 400$ MHz is the lower frequency limit of CHIME~\cite{CHIMEFRB:2021srp}. The spectral morphology of FRB signals encodes both the intrinsic plasma conditions at the emission site and propagation effects along the line of sight. The higher spectral indices of repeaters are consistent with observations, such as the detection of bursts from FRBs 121102~\cite{Law2017} and 180916~\cite{Pastor-Marazuela:2020tii} in some frequency ranges but not others. Additionally, two bursts were detected from FRB 200428 by CHIME~\cite{CHIMEFRB:2020abu}, but only one by STARE2~\cite{Bochenek:2020zxn}, which observes a higher frequency range than CHIME. Similar trends were identified in~\cite{Sun:2024huw, Junior:2025tuf}, consistent with prior findings that bursts from repeaters are intrinsically narrower than those from non-repeaters~\cite{Scholz:2016rpt, CHIMEFRB:2019pgo, Fonseca:2020cdd, CHIMEFRB:2021srp, CHIMEFRB:2023myn}.

This trend appears not to be an instrumental artifact, as a systematic study of 700 bursts from FRB 20201124A using the FAST telescope also found repeaters to exhibit significantly narrower bandwidths~\cite{Zhou:2022nnh}. A plausible physical interpretation is that narrowband emission arises from coherent magnetospheric processes, where radiation is produced in dense plasma environments and is inherently band-limited by local plasma and magnetic field conditions.  By contrast, the broader spectra of non-repeating FRBs may reflect cataclysmic or shock-driven emission in more diffuse environments, producing broadband radiation. Further data, however, will be essential to definitively disentangle intrinsic from instrumental effects.

The model also identifies dispersion measure excess, defined as the total dispersion measure minus the Milky Way contribution, as a significant predictor for repeater likelihood. Since DM excess serves as a proxy for source distance and redshift, this suggests that detected repeaters tend to be closer than non-repeaters.
This observed correlation likely arises primarily from selection effects. Confirming a source as a repeater requires re-observing the same sky location multiple times, which is only practical for nearby, bright, or frequently active sources. Additionally, bursts from more distant sources suffer greater dispersion and scattering, making them harder to detect repeatedly. Indeed, some events classified as non-repeaters may represent only the most luminous bursts from sources that do repeat, but too rarely or faintly for follow-up detections.
However, we cannot entirely rule out a physical interpretation. If the correlation is not purely observational, it could indicate that repeating sources are intrinsically less energetic than non-repeaters. This would align with theoretical expectations: cataclysmic one-off events such as core-collapse supernovae or compact object mergers (neutron stars, black holes, white dwarfs) typically involve higher energy budgets than models of repeated emission from, for example, young hyperactive magnetars. At present, disentangling selection effects from genuine physical differences remains an open challenge.

Another key result of this work is the identification of several false positives, non-repeating sources that our sVAE model classifies as repeaters. A list of these sources is provided in Table~\ref{tab:false_positives}. These non-repeaters exhibit latent-space features that are highly consistent with those of known repeaters. Interestingly, two of these sources (FRB20181218C and FRB20190221A) have been independently flagged as repeater candidates in prior machine learning analyses~\citep{Luo:2022smj, Zhu-Ge:2022nkz, J_nior_2026}. This motivates continued monitoring of these sources to search for additional bursts. Confirmation of repeating behavior through follow-up observations would offer valuable insight into the boundary between apparently non-repeating and genuinely repeating FRBs.

Despite these promising findings, our analysis has several limitations. The results are affected by CHIME’s observational biases, including its restricted 400–800 MHz frequency coverage and declination-dependent sky sensitivity due to its transit design. In addition, our approach relies on derived features rather than raw time-series data, which may omit subtle information not captured in the pre-processed dataset. We conclude that sVAEs are a useful tool to examine the underlying structure of FRB signals. Furthermore, we suggest that data from other FRB catalogs could be incorporated to improve the latent representation of FRBs, provided that the data can be homogenized to avoid artificially introducing experiment-dependent categorical substructures.

\section{Acknowledgements}

We would like to acknowledge Dongzi Li and Allen Foster for their guidance on domain-specific FRB topics. In addition, we would like to thank the University of Pittsburgh for hosting the 2025 Phenomenology Symposium, where an initial version of this paper was presented. AP acknowledges support from the Department of Energy (DOE) under Award Number DE-SC0007968.

\bibliographystyle{unsrt}
\bibliography{ref}

\section{Appendix}

\subsection{Libraries and Reproducibility}
\label{sec:reproducibility}

All experiments were implemented using Python 3.12.3 with the following libraries: PyTorch 2.6.0 for deep learning model implementation~\citep{torchpaper}, Optuna for hyperparameter optimization~\citep{optunapaper}, scikit-learn for evaluation metrics~\citep{sklearnpaper}, NumPy and Pandas for data manipulation, UMAP-learn and scikit-learn's t-SNE and Isomap for dimensionality reduction~\citep{umappaper, sklearnpaper}, and Matplotlib and Seaborn for visualization.

Model training and evaluation were conducted on an NVIDIA A10G GPU with CUDA support. All random seeds were fixed across NumPy, PyTorch, and Python's random module to ensure reproducible results. The stratified cross-validation splits were deterministic to enable exact replication of our findings.

All data and code are available and reproducible at
\href{https://github.com/rtenacity/frb-analysis}{GitHub}
and on Zenodo \citep{code_zenodo_doi}.

\end{document}